\documentclass[entropy,article,accept,pdftex,moreauthors]{Definitions/mdpi} 
\usepackage{amsmath} % For mathematical environments
\usepackage{amssymb} % For additional symbols
\usepackage{graphicx} % For including images
\usepackage{mathtools} % Extends amsmath and provides \mathclap

\firstpage{1} 
\pubvolume{1}
\issuenum{1}
\articlenumber{0}
\pubyear{2025}
\copyrightyear{2025}
\externaleditor{Dennis Dieks}
\datereceived{24 December 2024} 
\daterevised{23 January 2025} % Comment out if no revised date
\dateaccepted{29 January 2025} 
\datepublished{ } 
\hreflink{https://doi.org/} % If needed use \linebreak

\Title{Extending the QMM Framework to the Strong and Weak~Interactions}

\TitleCitation{Extending the QMM Framework to the Strong and Weak Interactions}

\Author{Florian Neukart %MDPI: Please carefully check the accuracy of names and affiliations. FN: the names are correct
 $^{1,2,}$*\orcidA{}, Eike Marx $^{2}$ and Valerii Vinokur $^{2}$ }

\AuthorNames{Florian Neukart, Eike Marx and Valerii Vinokur}

\AuthorCitation{Neukart, F.; %MDPI: Please check all author names carefully. FN: correct
 Marx, E.; Vinokur, V.}

\address{%
$^{1}$ \quad Leiden Institute of Advanced Computer Science, Leiden University, Gorlaeus Gebouw-BE-Vleugel, Einsteinweg 55, 2333 Leiden, The Netherlands\\
$^{2}$ \quad Terra Quantum AG, Kornhausstrasse 25, 9000 St. Gallen, Switzerland; 
eike@terraquantum.swiss (E.M.); vv@terraquantum.swiss (V.V.) %MDPI: We added the email addresses here according to those submitted online at susy.mdpi.com. Please check if they are correct. FN: these are correct, thank you
}

\corres{Correspondence: f.neukart@liacs.leidenuniv.nl}

\abstract{
We extend the Quantum Memory Matrix (QMM) framework, originally developed to reconcile quantum mechanics and general relativity by treating space--time as a dynamic information reservoir, to incorporate the full suite of Standard Model gauge interactions. In this discretized, Planck-scale formulation, each space--time cell possesses a finite-dimensional Hilbert space that acts as a local memory, or \emph{quantum imprint %MDPI: Please confirm if the italics is unnecessary and can be removed. Same as the following. Please check throughout the manuscript. FN: this helps with the newly introduced terms
}, for matter and gauge field configurations. We focus on embedding non-Abelian SU(3)\(_\mathrm{c}\) (quantum chromodynamics) and SU(2)\(_\mathrm{L}\)\(\times\)U(1)\(_Y\) (electroweak interactions) into QMM by constructing gauge-invariant imprint operators for quarks, gluons, electroweak bosons, and the Higgs mechanism. This unified approach naturally enforces unitarity by allowing black hole horizons, or any high-curvature region, to store and later retrieve quantum information about color and electroweak charges, thereby preserving subtle non-thermal correlations in evaporation processes. Moreover, the discretized nature of QMM imposes a Planck-scale cutoff, potentially taming UV divergences and modifying running couplings at trans-Planckian energies. We outline major challenges, such as the precise formulation of non-Abelian imprint operators and the integration of QMM with loop quantum gravity, as well as possible observational strategies—ranging from rare decay channels to primordial black hole evaporation spectra—that could provide indirect probes of this discrete, memory-based view of quantum gravity and the Standard Model. 
}


\begin{document}
%%%%%%%%%%%%%%%%%%%%%%%%%%%%%%%%%%%%%%%%%%

\section{Introduction}\label{sec:intro}

The unification of quantum mechanics (QM) with the full set of fundamental interactions remains one of the most ambitious goals of modern theoretical physics. Tremendous progress has been made in describing electromagnetic, weak, and strong forces within the Standard Model (SM) framework~\cite{weinberg1995quantum, peskin1995introduction}, and early successes in quantum gravity research—such as loop quantum gravity~\cite{rovelli1998loop, ashtekar2004background} and string theory~\cite{polchinski1998, maldacena1998}—have offered valuable insights into the nature of space--time at the Planck scale. Nevertheless, a fully unified, experimentally testable theory that elegantly incorporates all %MDPI: Please confirm if the italics is unnecessary and can be removed. Same as the following. Please check throughout the manuscript. FN: this can be removed.
 known interactions remains~elusive.

A central difficulty lies in reconciling the apparent tension between quantum unitarity and the classical description of horizons and singularities in general relativity (GR). The black hole information paradox~\cite{hawking1976, hawking1975, unruh1976notes, preskill1992, giddings2007black}, in particular, epitomizes this conflict by suggesting that information about matter falling into a black hole might be irretrievably lost if Hawking radiation is purely thermal. Such a loss would violate unitarity, a cornerstone principle of quantum mechanics. Although various resolution proposals (firewalls~\cite{almheiri2013}, fuzzballs~\cite{mathur2005}, holography~\cite{tHooft1993Dimensional, susskind1995}, ER = EPR %MDPI: Please confirm if it is a whole. If not, please add space before and after =, e.g., ER = EPR. FN: generally, no preference, but I added blanks before and after
~\cite{maldacena2013cool}, etc.%MDPI: We removed the italics of Latin expression. Please confirm. FN: OK
) have advanced our understanding of black holes, no consensus has emerged on a definitive solution that accommodates the entire Standard Model (including SU(3)\(_\mathrm{c}\) for color and SU(2)\(_\mathrm{L}\)\(\times\)U(1)\(_Y\) for electroweak interactions) alongside quantum gravity.

\subsection{Background and Motivation}

In prior work, the \emph{Quantum Memory Matrix} (QMM) approach was proposed to address the black hole information paradox by interpreting space--time as a dynamic quantum information reservoir~\cite{neukart2024quantum}. This perspective treats each Planck-scale region of space--time as a finite-dimensional Hilbert space capable of “storing” local quantum data via \emph{quantum imprints}. By integrating QMM with gravitational degrees of freedom, it was shown that information from infalling matter could be preserved in and later retrieved from the structure of space--time, thus allowing black hole evaporation to remain unitary without necessitating exotic boundary conditions, wormhole-based nonlocality, or event \mbox{horizon firewalls.}

Subsequent extensions of the QMM framework to \emph{electromagnetism}~(U(1) gauge theory) demonstrated that local gauge invariance can be upheld in a discretized space--time by constructing suitable \emph{imprint operators} from electromagnetic field strengths and currents. This development showed that QMM can embed quantum information about electric charge and photon modes at Planck-scale cells, opening the path to a more general inclusion of Standard Model interactions.

\subsection{Aim and Scope of This Paper}

The primary purpose of this work is to extend the QMM hypothesis further by \mbox{incorporating}:
\begin{enumerate}
    \item The strong interaction%MDPI: Please confirm if the bold is unnecessary and can be removed. Same as the following. Please check throughout the manuscript. FN: can be removed
, governed by non-Abelian SU(3)\(_\mathrm{c}\) gauge fields (QCD), including color confinement and gluon self-interactions.
    \item The weak interaction, governed by SU(2)\(_\mathrm{L}\)\(\times\)U(1)\(_Y\), encompassing chiral symmetry breaking, flavor mixing, and the Higgs mechanism.
\end{enumerate}

In doing so, we aim to build on the local, discretized QMM structure to show how color and electroweak charges, along with their associated gauge bosons, can be imprinted in Planck-scale space--time quanta. Specifically, we will:

\begin{itemize}
    \item Outline how the strong force (quarks, gluons, color flux tubes) can be captured by imprint operators that remain \emph{locally gauge-invariant} in a discretized SU(3) scenario.
    \item Demonstrate how the weak isospin and hypercharge interactions (SU(2)\(_\mathrm{L}\)\(\times\)U(1)\(_Y\)) couple to QMM cells and how spontaneous symmetry breaking (including the Higgs field) fits into the QMM picture.
    \item Present a unifying, \emph{QMM-based} Hamiltonian that combines gravity (in discrete form), QCD, and the electroweak interactions under the same Planck-scale “memory” \mbox{framework}.
    \item Discuss how colored and electroweak-charged matter falling into a black hole leaves \emph{quantum imprints} that preserve unitarity in the Hawking evaporation process.
    \item Propose potential observational consequences, including subtle effects on hadronic collisions, neutrino oscillations, rare decay processes, and astrophysical black \mbox{hole evaporation}.
\end{itemize}

\subsection{Structure of This Paper}

Following this introduction, we provide, in Section~\ref{sec:review}, a brief overview of the Quantum Memory Matrix hypothesis and its previous applications to gravity and electromagnetism. In Section~\ref{sec:strong}, we incorporate SU(3)\(_\mathrm{c}\) into QMM, focusing on how color confinement and quark/gluon degrees of freedom naturally fit into space--time quanta. Section~\ref{sec:weak} discusses the extension to the weak sector, covering SU(2)\(_\mathrm{L}\)\(\times\)U(1)\(_Y\), spontaneous symmetry breaking, and flavor transitions. In Section~\ref{sec:unifiedHamiltonian}, we synthesize these elements into a unified QMM Hamiltonian that includes gravitational, strong, and electroweak fields. Section~\ref{sec:applications} explores implications for black hole evaporation, baryogenesis, and potential observational or experimental signatures. Section~\ref{sec:comparison} compares QMM with other unifying frameworks—such as AdS/CFT, ER=EPR, loop quantum gravity, and minimal length theories—highlighting QMM’s local and covariant advantages. Section~\ref{sec:challenges} then addresses open technical and conceptual issues, including precise non-Abelian imprint construction and the renormalization group flow at trans-Planckian scales. Finally, in Section~\ref{sec:conclusion}, we summarize our findings and outline future research directions aimed at achieving a fully discrete, information-centric model of the Standard Model plus gravity.

By solidifying how the strong and weak interactions can be encoded in a Planck-scale quantum memory structure, this paper aspires to move one step closer to a consistent theory of quantum gravity and gauge interactions in which no information is ever truly lost, even in the most extreme conditions found at black hole horizons.

\medskip

\subsection{Additional Remarks Addressing Foundational Concerns} %MDPI: Please confirm if this paragraph' format is correct and should be kept. FN: made a subsection 
\emph{Before} proceeding, we acknowledge two fundamental issues raised by critics of discrete unification schemes---notably the question of \textit{informational energy} and the concern of \textit{Lorentz invariance} in a fixed-cell picture. We emphasize:
\begin{enumerate}
    \item Informational Energy Budget. The QMM cell’s Hilbert space, while finite-dimensional, does not necessarily imply a large or infinite extra energy reservoir. Rather, these discrete internal states function much like spin or field modes regulated by quantum principles. As in standard field theory, these vacuum-like modes can be renormalized, and any net “informational energy” may be negligible or integrated into the vacuum sector. A fully quantitative treatment remains an open question (see Section~\ref{sec:challenges}), but we do not require a radical new form of matter or dark energy. 
    \item Lorentz Invariance and the Preferred Frame Issue. While a lattice or cell structure can look like a rest-frame artifact, many approaches to discrete quantum gravity (e.g., causal sets, loop quantum gravity, spin foams) are also built on fundamental discreteness yet recover effective Lorentz symmetry at continuum scales. Similarly, QMM is posited as a Planck-scale regularization that \emph{emerges} as a smooth, relativistically covariant manifold at energies far below $M_{\mathrm{P}}$. We discuss these conceptual parallels further in Section~\ref{sec:comparison}.
\end{enumerate}

Finally, we stress that although our approach aims at unification, it does so in a measured fashion, reminiscent of the caution that both Feynman and Glashow voiced about ambitious "theory of everything" proposals. In particular, the QMM does not presume a near-term empirical verification at the Planck scale, but rather provides a \emph{discrete information} viewpoint that may be tested indirectly through black hole phenomenology, rare decay channels, or cosmological constraints. We now turn to a more detailed review of the QMM framework as it applies first to gravity and electromagnetism, then extend it to strong and weak interactions.

\section{Review of the QMM Framework}
\label{sec:review}

The QMM framework was originally introduced to address the longstanding tension between quantum unitarity and the classical description of horizons in general \mbox{relativity~\cite{neukart2024quantum}}. It interprets space--time not merely as a geometric background but as a quantum information \emph{reservoir}, discretized at the Planck scale. In this section, we summarize the key principles and mathematical underpinnings of the QMM. We begin by describing how space--time is discretized into Planck-scale cells, each equipped with a finite-dimensional Hilbert space. We then recapitulate the concept of \emph{quantum imprints}, which store local quantum information, ensuring that unitarity and locality are preserved. Finally, we highlight how the QMM has so far been successfully applied to reconciling black hole evaporation with unitarity in a purely gravitational setting, as well as in the presence of electromagnetic interactions.

\subsection{Discretization of Space--Time and Finite Hilbert Spaces}
\label{subsec:QMM_discretization}

A primary postulate of the QMM approach is that the continuum description of \mbox{space--time} ceases to be valid at the Planck scale, $l_{\mathrm{P}} \approx 1.616\times 10^{-35}\,\mathrm{m}$. Instead, space--time is modeled as a \emph{discrete} set of fundamental cells, each with volume on the order of $l_{\mathrm{P}}^3$ (plus the Planck time $t_{\mathrm{P}}$ when considering temporal discretization). These \emph{Planck cells}, indexed by $x$ in a set $\mathcal{X}$, collectively form the foundation of the QMM picture%MDPI: References should be numbered in order of appearance. we rearranged all the references to appear in numerical order. Please check. FN: these are in order, as they have appeared in previous sections
~\cite{rovelli1998loop, ashtekar2004background, hossenfelder2013minimal}.

\begin{figure}[H]

\includegraphics[width=0.725\textwidth]{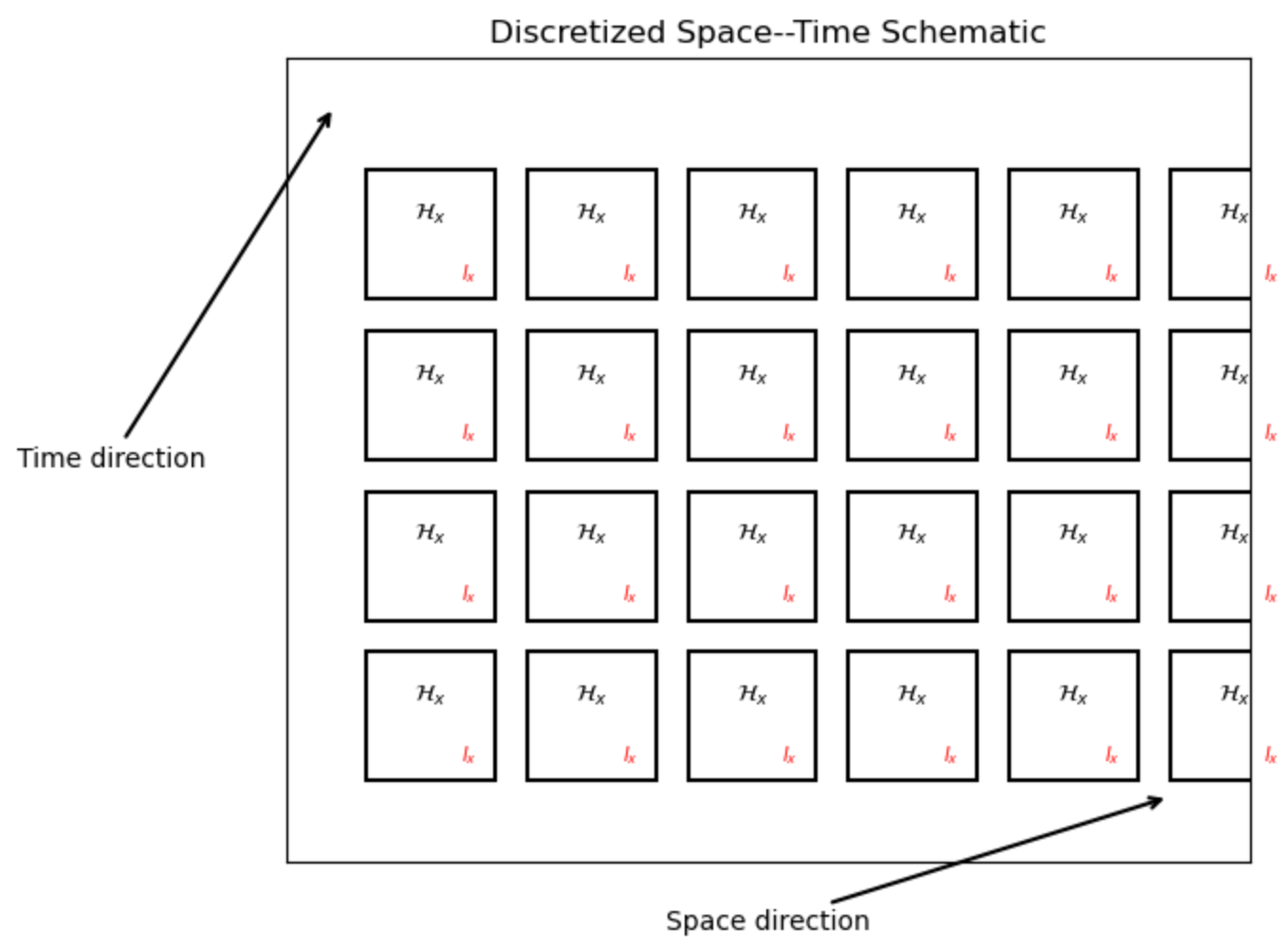}
\vspace{-2.0pt}
\caption{A 2D representation of discretized space--time at the Planck scale, where each cell is associated with a finite-dimensional Hilbert space \(\mathcal{H}_x\). The label \(\mathcal{H}_x\) indicates the local degrees of freedom stored in that cell, while small icons (or “\(I_x\)” marks) represent possible imprint operators recording matter or gauge interactions. Time and space directions are only schematic here.}
\label{fig:discretized_spacetime}
\end{figure}

Each cell $x \in \mathcal{X}$ carries a finite-dimensional Hilbert space $\mathcal{H}_x$ whose dimension is large enough to encode local geometric and field-theoretic degrees of freedom, and yet remains bounded to impose a natural ultraviolet (UV) cutoff. The total QMM Hilbert space is a tensor product: %MDPI: Please check throughout the manuscript to ensure the formats (italic, bold, sub/superscript, etc.) of the variables and parameters are consistent and revise if necessary. FN: thank you, all is correct
\begin{equation}\label{eq:H_QMM}
    \mathcal{H}_{\mathrm{QMM}} \;=\; \bigotimes_{x \in \mathcal{X}} \;\mathcal{H}_x.
\end{equation}
In such a discretized scheme, operators corresponding to geometric observables, e.g., quantized area or volume operators, gain discrete spectra, akin to loop quantum gravity approaches~\cite{rovelli1996black, ashtekar2005}. Unlike purely geometric formalisms, however, QMM interprets each cell $\mathcal{H}_x$ as an \emph{active memory unit}, capable of storing quantum-state data of matter or gauge fields that locally interact with it. A schematic depiction of these Planck-scale cells, each labeled by a local Hilbert space \(\mathcal{H}_x\), is shown in Figure~\ref{fig:discretized_spacetime}.

\vspace{-5.0pt}

Because each cell is associated with a \emph{finite} dimensional space, integrals over momenta or field configurations are replaced by sums over discrete modes, naturally regularizing the high-energy regime. This discretization not only prevents the infinities commonly encountered in continuum quantum field theory but also provides a physically motivated cutoff at the Planck scale~\cite{hossenfelder2013minimal}. As discussed in Section~\ref{sec:comparison}, the question of how such cells reconcile with macroscopic Lorentz invariance and overall energy content in cosmology remains an open area of investigation.

\subsection{Quantum Imprints and Local Encoding of Information}
\label{subsec:QMM_imprints}

The key innovation of the QMM is the notion of \emph{quantum imprints}~\cite{neukart2024quantum}, which are the local records that store quantum information about field interactions or particle states passing through a given Planck cell. Each cell $x$ thus becomes an interactive element in spacetime, contributing to the dynamical evolution of the entire system rather than a mere geometric backdrop.

\subsubsection{Imprint Operators %MDPI: We changed to the paragraph into subsubsection title. Please confirm the modification. Same as the following highlights. FN: this is OK.
} 
Formally, the imprinting process is captured by imprint operators $\hat{I}_x$, which act on the tensor product space
\[
    \mathcal{H}_\text{fields} \;\otimes\; \mathcal{H}_x.
\]
If a field operator $\hat{\phi}(x)$ (e.g., a scalar, fermionic, or gauge field) interacts at cell $x$, the local Hamiltonian takes the schematic form
\begin{equation}
    \hat{H}_{\text{int}} \;=\; \sum_{x \in \mathcal{X}} \Bigl(\hat{\phi}(x)\,\otimes\,\hat{I}_x^\dagger + \hat{\phi}^\dagger(x)\,\otimes\,\hat{I}_x\Bigr),
\end{equation}
reflecting a bidirectional flow of quantum information: the cell “records” data about the field, and the field may in turn be affected by the cell's stored imprint state. Because these interactions appear in a global Hamiltonian that is Hermitian, the total evolution is unitary, and no net information is lost.

\subsubsection{Locality and Causality}
Interactions in QMM remain strictly local, confined to individual cells. This ensures that no superluminal signaling or acausal effects arise. Cells only exchange information with nearest neighbors, if at all, via carefully constrained interactions consistent with relativistic causality.

\subsubsection{Information Retrieval Mechanism}
Once an imprint is stored in a cell’s Hilbert space, it can be retrieved by future interactions at that cell. Notably, black hole horizons exemplify how inward-falling quantum states imprint on the horizon cells. As Hawking-like radiation emerges, it interacts again with these cells, gradually extracting the imprinted data in subtle (often non-thermal) correlations. Hence, QMM resolves the paradox of “information loss” by embedding the missing data in local degrees of freedom that rejoin outgoing radiation in a unitary process~\cite{hawking1976, hawking1975, neukart2024quantum}.

\subsection{Previous Results: QMM with Gravity and Electromagnetism}
\label{subsec:QMM_grav_em}

\subsubsection{Gravity}
The first major application of QMM was to pure gravitational systems, particularly black holes~\cite{neukart2024quantum}. By discretizing spacetime around the horizon, QMM showed how to store and retrieve information from the horizon cells, allowing for a globally unitary description of black hole formation and evaporation. The equivalence principle and smooth horizon structure remained intact because the QMM’s imprinting occurs at Planck scales, leaving semiclassical geometry unaltered at macroscopic distances.

\subsubsection{Electromagnetism}
Incorporating U(1) gauge fields into QMM introduced gauge-invariant imprint operators built from field strengths $F_{\mu\nu}$ and charged matter currents~\cite{peskin1995introduction, neukart2024quantum}. This extension demonstrated that local electromagnetic data (charge, photon states) could imprint into Planck cells without breaking gauge invariance. Charged black hole (Reissner--Nordström) evaporation was then modeled in QMM, preserving unitarity by correlating the U(1) field with the local memory of spacetime quanta.

\subsubsection{Rationale for SU(3)$_\mathrm{c}$ and SU(2)$_\mathrm{L}$ $\times$ U(1)$_{\mathrm{Y}}$}
While these initial successes established the viability of the QMM as a unitarity-preserving framework for gravity and Abelian gauge fields, the Standard Model also contains non-Abelian gauge sectors—namely SU(3)\(_\mathrm{c}\) for QCD and SU(2)\(_\mathrm{L}\)\(\times\)U(1)\(_Y\) for the electroweak interactions. Incorporating these into the QMM is essential for describing quarks, gluons, electroweak bosons, flavor transitions, and related phenomena, especially under extreme gravitational conditions such as black hole collapse or the early universe. 

\medskip
 %MDPI: Please confirm if this paragraph' format is correct and should be kept. FN: This is correct, but we can remove the header

In parallel with QMM-based perspectives, there have been a variety of newer models suggesting that black holes behave like \emph{hollow massive shells} subject to strong quantum fluctuations~\cite{ShellModel2014,ShellModel2023,ShellModel2022,ShellModel2024}. These approaches propose that the Schwarzschild radius is not a true horizon but an apparent one, where collapsing matter condenses into a shell-like structure. While the QMM picture places local memory degrees of freedom at or near the horizon cells, the shell-based scenario views the black hole boundary as a quantum region governed by a Schr\"odinger-like equation. In principle, both frameworks aim to solve the information paradox by modifying the classical notion of an empty black hole interior. A detailed reconciliation—whether QMM cells can be viewed as a discrete realization of this shell boundary, or whether hollow-shell solutions can embed QMM-like local unitarity—remains an open question. We note that each approach highlights the possibility that the “classical horizon” might be replaced by a quantum region with additional degrees of freedom, consistent with the general idea that black holes are not truly featureless gravitational objects.

\medskip
Thus, the next sections focus on how SU(3)\(_\mathrm{c}\) and SU(2)\(_\mathrm{L}\)\(\times\)U(1)\(_Y\) integrate into QMM in a way that preserves unitarity and local gauge invariance, building toward a unified discrete framework for the entire Standard Model plus gravity.

\section{Incorporating the Strong Interaction, \texorpdfstring{SU(3)\(_\mathrm{c}\)}{SU(3)c}}
\label{sec:strong}

Having established how the QMM framework accommodates quantum gravity and Abelian gauge fields~\cite{neukart2024quantum}, we now extend QMM to the non-Abelian sector of the Standard Model: the \emph{strong interaction} governed by SU(3)\(_\mathrm{c}\). Quantum Chromodynamics (QCD) underlies hadron formation, color confinement, and gluon self-interactions~\cite{peskin1995introduction, weinberg1995quantum}, making it significantly more intricate than Abelian electromagnetism. Here, we demonstrate how the QCD color degrees of freedom---quarks, gluons, and their local color charges---can be embedded in the Planck-scale QMM cells via \emph{color imprint operators} that preserve the local SU(3) gauge invariance, while still integrating with the gravitational and electromagnetic aspects of QMM.

\subsection{Gauge Fields and Color Charge in the QMM Framework}
\label{subsec:strong_gauge_fields}

In continuum QCD, the gluon fields \(G_{\mu}^a(x)\), where \(a = 1,\dots,8\), mediate interactions among quarks, each carrying a color index (red, green, blue). The SU(3) gauge symmetry is manifested in the covariant derivative
\begin{equation}\label{eq:qcd_covariant_derivative}
    D_{\mu} \;=\; \partial_{\mu} \,\mathbf{1} \;-\; i\,g_s\, T^a\, G_{\mu}^a,
\end{equation}
where \(g_s\) is the strong coupling constant, and \(T^a\) are the SU(3) generators in the appropriate representation~\cite{peskin1995introduction, wilson1974}. Implementing this structure in a discretized QMM setting requires:

\begin{enumerate}
    \item Local SU(3) Invariance: Each Planck cell \(x\) must interact with quark and gluon fields in a manner consistent with \(\mathrm{SU}(3)\) gauge transformations. The QMM imprint operators \(\hat{I}^{(\mathrm{color})}_x\) must be strictly gauge-invariant or gauge-covariant such that physical observables remain invariant under local \(\mathrm{SU}(3)\) transformations.
    \item Finite-Dimensional Hilbert Spaces: While continuum QCD allows infinitely many field modes, QMM discretization ensures that each cell \(x\) has a finite-dimensional space. We choose a dimension sufficient to encode local color states (e.g., quark triplets and gluon octets) while imposing a Planck-scale UV cutoff.
\end{enumerate}

One practical approach for including gluon fields is to employ link variables \(\hat{U}_{x,\mu} \approx \exp\bigl[i\,a\, g_s\, T^a\, G_{\mu}^a(x)\bigr]\) connecting neighboring cells, akin to lattice QCD~\cite{wilson1974}, but endowed with QMM’s memory functionality. Alternatively, one may introduce \(\hat{G}_{\mu}^a(x)\) directly in each cell, provided the color imprint operators retain the required local invariance. We highlight that reconciling such a discretization with potential large-scale Lorentz invariance and energy constraints is discussed later (see Section~\ref{sec:comparison}).

\subsection{Confinement and the QMM Interpretation}
\label{subsec:strong_confinement}

A hallmark of QCD is \emph{color confinement}: the empirical fact that color-charged particles (quarks and gluons) cannot be isolated at macroscopic distances, forming instead color-neutral hadrons. In lattice QCD, this is often conceptualized via \emph{flux tubes} or Wilson loops~\cite{wilson1974}, which indicate a linearly growing potential between static color charges.

\begin{figure}[H]
\vspace{-3.0pt}
\includegraphics[width=0.9\textwidth]{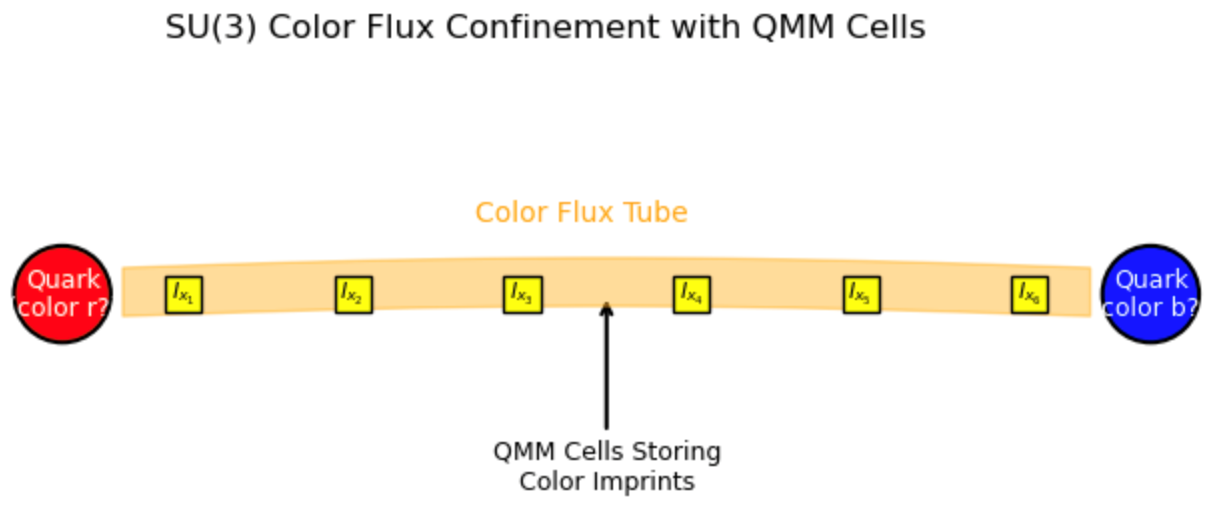}
\vspace{-2.0pt}
\caption{A conceptual %MDPI: Please cite the figure in main text and ensure that the first citation of each figure appears in numerical order. FN: it is cited above
 diagram of SU(3) color flux confinement in QMM. Two quarks (red and blue) are connected by a color flux tube. Discrete QMM cells along the line (represented by small squares) store local color imprints \(\{I_{x_1}, I_{x_2}, \dots\}\) that enforce confinement.}
\label{fig:color_flux_confinement}
\end{figure}

From the QMM standpoint, we propose that confinement naturally arises as color flux lines become \emph{imprinted} in the Planck cells along the line connecting quarks. This notion of color flux tubes stored by QMM cells is illustrated in Figure~\ref{fig:color_flux_confinement}, showing two quarks connected by a flux tube with local imprint sites. As two color charges attempt to separate, more and more QMM cells between them store color-imprint information, effectively creating a “string” of color flux. Breaking the flux tube costs energy to produce quark--antiquark pairs. Thus, color-neutral hadrons correspond to net cancellations of these local color imprints within finite volumes of QMM cells.

Because each cell can hold only a finite number of color-imprint states, the framework suggests a discretized mechanism for flux tube formation. At distances well above the Planck scale, this reproduces standard continuum confinement, while near or beyond trans-Planckian energies, the QMM memory limit might saturate or modify the \mbox{confining potential.}

\subsection{Quarks, Gluons, and Local Imprint Operators}
\label{subsec:quarks_gluons_imprints}

\subsubsection{Quark Field Coupling}
In continuum QCD, a quark field \(\hat{\psi}_\alpha^i(x)\) carries both spinor (\(\alpha\)) and color (\(i = r,g,b\)) indices. In QMM, for each cell \(x\), we maintain a local quark operator \(\hat{\psi}_\alpha^i(x)\) interacting with the QMM cell’s Hilbert space \(\mathcal{H}_x\) via imprint operators

\begin{equation}\label{eq:color_imprint_quark}
    \hat{H}_{\mathrm{int}}^{(\mathrm{quark})} 
    \;=\; \sum_{x \in \mathcal{X}} \Bigl[
        \hat{\overline{\psi}}^i_\alpha(x)\, \gamma^\mu \,T^a_{ij}\,\hat{\psi}^j_\alpha(x)\;\hat{G}_\mu^a(x) 
        \;\otimes\; \hat{I}_x^{(\mathrm{color})}\Bigr] \;+\;\mathrm{h.c.},
\end{equation}
where \(T^a\) are the SU(3) generators in the fundamental representation, and \(\hat{G}_\mu^a(x)\) is the discrete gluon operator. The factor \(\hat{I}_x^{(\mathrm{color})}\) acts on \(\mathcal{H}_x\), encoding the local color interaction at that cell.

\subsubsection{Gluon Self-Interaction}
Unlike QED, gluons themselves carry color charge. The gluon field \(\hat{G}_\mu^a\) thus has nontrivial self-couplings:
\begin{equation}\label{eq:gluon_self_coupling}
    f^{abc}\, \hat{G}_\mu^a \,\hat{G}_\nu^b \,\hat{G}_\rho^c,
\end{equation}
where \(f^{abc}\) are the SU(3) structure constants~\cite{peskin1995introduction}. In QMM, each cell’s imprint operator \(\hat{I}_x^{(\mathrm{color})}\) must be constructed to maintain gauge invariance under these non-Abelian self-interactions. One possible structure is
\begin{equation}\label{eq:color_imprint_op}
    \hat{I}_x^{(\mathrm{color})} \;=\; \lambda_0\, \hat{F}_{\mu\nu}^a(x)\,\hat{F}^{a,\mu\nu}(x) 
    \;+\;\lambda_1\, f^{abc}\,\hat{G}_\mu^a(x)\,\hat{G}_\nu^b(x)\,\ldots 
    \;+\;\cdots,
\end{equation}
where \(\hat{F}_{\mu\nu}^a\) is the non-Abelian field strength at cell \(x\), and the ellipses represent higher-order terms. The imprint operator must transform covariantly (or be explicitly invariant) under local SU(3) transformations.

\subsubsection{Unitarity and Local Gauge Invariance}
As in the electromagnetic case, the total Hamiltonian 
\[
    \hat{H}_{\mathrm{QCD}} + \hat{H}_{\mathrm{QMM}} + \hat{H}_{\mathrm{int}}^{(\mathrm{quark, gluon})}
\]
must be Hermitian to ensure global unitarity of the combined system, including QMM degrees of freedom. Local gauge transformations that rotate quark color indices or gluon fields at each cell leave physical observables invariant, preserving the fundamental SU(3) symmetry of QCD.

\subsection{Hadronization, Non-Perturbative Effects, and Planck-Scale Cutoff}
\label{subsec:hadronization}

Below the confinement scale $\Lambda_\mathrm{QCD}\sim \mathcal{O}(200\,\mathrm{MeV})$, hadrons form as color-neutral bound states. Non-perturbative phenomena such as chiral symmetry breaking and meson/baryon formation are notoriously challenging in continuum field theory. However, in QMM’s discretized view, two conceptual features emerge:

\begin{enumerate}
    \item Local Color Neutralization: As quarks and gluons traverse successive Planck cells, color flux lines are “recorded” via imprint operators. Hadrons correspond to quark/gluon configurations whose net imprint over a finite region of QMM cells sums to a color singlet.
    \item Ultraviolet Completion: Because each cell’s Hilbert space dimension is finite, a natural cutoff arises near the Planck scale. In principle, this could unify non-perturbative QCD with quantum gravity without the usual continuum field-theoretic divergences.
\end{enumerate}

Moreover, at energies near or exceeding the Planck scale, the QMM memory limit could saturate, altering hadronization or parton showering beyond standard QCD expectations. While such regimes lie far beyond present experiments, they are crucial to understanding black hole formation by QCD matter or early-universe epochs where strong interaction and gravity intermix~\cite{weinberg1995quantum}. Having demonstrated how SU(3)\(_\mathrm{c}\) can be incorporated into the QMM’s discretized, local-memory framework, we now turn to the \emph{electroweak} sector—SU(2)\(_\mathrm{L}\)\(\times\)U(1)\(_Y\)—to complete the Standard Model gauge structure. In the next section, we examine how weak isospin, hypercharge, and spontaneous symmetry breaking also fit into QMM, paving the way for a unified treatment of black hole evaporation and other high-energy phenomena.

\section{Extending QMM to the Weak Interaction (\texorpdfstring{SU(2)\(_\mathrm{L}\)\(\times\)U(1)\(_Y\)}{SU(2)L x U(1)Y})}
\label{sec:weak}

The electroweak sector of the Standard Model (SM) is governed by the non-Abelian SU(2)\(_\mathrm{L}\) gauge group and an additional U(1)\(_Y\) symmetry, later spontaneously broken down to the electromagnetic U(1)\(_{\mathrm{em}}\) via the Higgs mechanism~\cite{peskin1995introduction,weinberg1995quantum}. In the QMM framework, we must ensure that both \emph{local} SU(2)\(_\mathrm{L}\) and \emph{local} U(1)\(_Y\) invariance are preserved at each Planck-scale cell and that the resulting massive gauge bosons (W and Z) and massless photon emerge naturally after spontaneous symmetry breaking. Additionally, QMM must accommodate flavor transitions, such as neutrino oscillations and quark mixing, which introduce subtle quantum phases. Below, we outline how SU(2)\(_\mathrm{L}\)\(\times\)U(1)\(_Y\) can be embedded in QMM, focusing on gauge fields, imprint operators, and the Higgs field.

\subsection{Weak Isospin, Hypercharge, and the QMM Cell Structure}
\label{subsec:weak_isospin_hypercharge}

In continuum electroweak theory, the gauge fields are:
\begin{itemize}
    \item \(\hat{W}_{\mu}^i(x)\), for \(i=1,2,3\), associated with SU(2)\(_\mathrm{L}\),
    \item \(\hat{B}_{\mu}(x)\), associated with U(1)\(_Y\).
\end{itemize}

Left-handed fermions (quarks and leptons) transform as SU(2)\(_\mathrm{L}\) doublets with weak isospin \(T = \tfrac12\), while their hypercharge \(Y\) dictates how they interact with \(\hat{B}_{\mu}(x)\). 

\subsubsection{Discretized Gauge Fields}
In the QMM discretization, each Planck cell \(x \in \mathcal{X}\) is extended with local operators \(\hat{W}_{\mu}^i(x)\) and \(\hat{B}_{\mu}(x)\). Alternatively, one may adopt link variables \(\hat{U}_{x,\mu}^{(\mathrm{SU(2)}})\approx \exp\!\bigl[i\,g\, \tau^i\,\hat{W}_{\mu}^i(x)\bigr]\) and \(\hat{U}_{x,\mu}^{(\mathrm{U(1)}})\approx \exp\!\bigl[i\,g'\,\hat{B}_{\mu}(x)\bigr]\), in analogy to lattice formulations. Here, \(g\) and \(g'\) are the usual SU(2)\(_\mathrm{L}\) and U(1)\(_Y\) coupling constants, while \(\tau^i\) are the SU(2) generators (Pauli matrices up to a factor).

\subsubsection{Local Imprint Operators}
To preserve gauge invariance, the QMM imprint operators \(\hat{I}_x^{(\mathrm{weak})}\) must transform covariantly (or remain strictly invariant) under SU(2)\(_\mathrm{L}\)\(\times\)U(1)\(_Y\). For instance, one can build imprint operators from the non-Abelian field strength tensors
\begin{equation}
    \hat{W}_{\mu\nu}^i \;=\; \partial_{\mu}\hat{W}_{\nu}^i \,-\,\partial_{\nu}\hat{W}_{\mu}^i 
    \;+\; g\,\epsilon^{i j k}\,\hat{W}_{\mu}^j\,\hat{W}_{\nu}^k,
\end{equation}
and from the Abelian-like hypercharge field strength \(\hat{B}_{\mu\nu} = \partial_{\mu}\hat{B}_{\nu} - \partial_{\nu}\hat{B}_{\mu}\). Such combinations can be inserted into local operators (as in lattice gauge theory~\cite{wilson1974, peskin1995introduction}):
\begin{equation}\label{eq:weak_imprint_op}
    \hat{I}_x^{(\mathrm{weak})} \;=\;
    \alpha_1\, \hat{W}_{\mu\nu}^i\,\hat{W}^{i,\mu\nu} 
    \;+\; \alpha_2\, \hat{B}_{\mu\nu}\,\hat{B}^{\mu\nu}
    \;+\;\alpha_3\,(\text{weak currents}) \;+\;\ldots
\end{equation}
acting on \(\mathcal{H}_x\). Here, \(\alpha_1,\alpha_2,\alpha_3\) specify coupling strengths to the QMM, and “weak currents” may include lepton or quark doublet operators \(\hat{\psi}_L\) in a gauge-invariant manner.

\subsection{Spontaneous Symmetry Breaking and the Higgs Mechanism}
\label{subsec:higgs_sbp}

A core feature of the electroweak theory is the Higgs mechanism: an SU(2)\(_\mathrm{L}\)\(\times\)U(1)\(_Y\) scalar doublet \(\Phi\) acquires a nonzero vacuum expectation value (vev), spontaneously breaking the gauge symmetry down to U(1)\(_{\mathrm{em}}\). The gauge bosons \(W^\pm\) and \(Z\) become massive, while the photon remains massless~\cite{peskin1995introduction, weinberg1995quantum}.

\subsubsection{Higgs Field in QMM}
In the QMM framework, the Higgs field \(\hat{\Phi}(x)\) appears as another operator at cell \(x\), storing or retrieving local quantum data through an interaction Hamiltonian term of \mbox{the form:}
\begin{equation}\label{eq:higgs_interaction}
    \hat{H}_{\text{Higgs--QMM}} 
    \;=\; \sum_{x}\Bigl[\hat{D}_\mu \hat{\Phi}^\dagger(x)\;\hat{D}^\mu \hat{\Phi}(x)\;\otimes\;\hat{I}_x^{(\mathrm{Higgs})}\Bigr],
\end{equation}
where \(\hat{D}_\mu\) is the electroweak covariant derivative (including \(\hat{W}_{\mu}^i, \hat{B}_{\mu}\)). The imprint operator \(\hat{I}_x^{(\mathrm{Higgs})}\) guarantees local interactions remain gauge-invariant and unitary. Once \(\hat{\Phi}\) acquires a vev, the QMM discretization distinguishes “broken” from “unbroken” cells, in the sense that the local vacuum state is shifted. The $W^\pm$ and $Z$ bosons naturally gain mass as excitations around this new vacuum configuration.

\subsubsection{Massive Gauge Bosons and U(1)\(_{\mathrm{em}}\)}
Post-symmetry breaking, the combinations \(\hat{W}^3_\mu\) and \(\hat{B}_\mu\) yield the massless photon field \(\hat{A}_\mu\) and the massive \(Z_\mu\). In QMM, each cell’s SU(2)\(\times\)U(1) operators are merely recast into “mixed” operators for \(A_\mu\) and \(Z_\mu\). The local imprint operators remain valid, simply reorganized to reflect the new basis in which one gauge boson is massless and the others carry mass. Hence, the QMM approach seamlessly accommodates the standard Higgs mechanism while leaving large-scale phenomenology unaltered.

\subsection{Flavor Transitions, Neutrino Mixing, and QMM}
\label{subsec:flavor_transitions}

Unlike QCD, which confines color charges, the electroweak sector allows quark flavor transitions (through charged-current interactions with \(W^\pm\)) and neutrino flavor oscillations (once neutrinos possess small masses). In the QMM framework, these phenomena can be seen as local \emph{flavor imprinting}:

\subsubsection{Charged Currents and CKM Matrix}
For quark doublets \(\hat{q}_L = (u_L, d_L)^T\), the QMM interaction Hamiltonian might include terms such as
\begin{equation}\label{eq:charged_current_CKM}
    \hat{H}_{\mathrm{cc}} 
    \;\sim\; \sum_x \Bigl[
       \hat{\overline{u}}_L(x)\;\gamma^\mu \;V_{\mathrm{CKM}}\;\hat{d}_L(x)\;\hat{W}_\mu^+(x) 
       \;\otimes\; \hat{I}_x^{(\mathrm{weak})}
       \;+\;\mathrm{h.c.}\Bigr],
\end{equation}
where \(V_{\mathrm{CKM}}\) is the Cabibbo--Kobayashi--Maskawa matrix mixing different generations. Within QMM, the local imprint operator \(\hat{I}_x^{(\mathrm{weak})}\) captures these flavor transitions in each cell, preserving overall unitarity while permitting quark flavor changes.

\subsubsection{Neutrino Mixing and PMNS Matrix}
Similarly, the lepton doublets \(\hat{l}_L = (\nu_L, e_L)^T\) may incorporate mixing among neutrino mass eigenstates through the Pontecorvo--Maki--Nakagawa--Sakata (PMNS) matrix~\cite{peskin1995introduction, weinberg1995quantum}. Each cell records a local imprint of these flavor transitions. Over large distances or times, the imprinting accumulates into neutrino oscillation patterns observed experimentally.

\subsubsection{Unitarity Preservation in Flavor Processes}
Because QMM uses a globally Hermitian Hamiltonian, the evolution of quark and lepton flavors across Planck cells remains manifestly unitary. No net flavor information is lost. Instead, QMM memory stores intermediate “snapshots” of the flavor state, later retrievable through subsequent interactions.

\subsection{Possible Experimental Signatures in Rare Decays}
\label{subsec:rare_decays}

Although direct Planck-scale tests are currently infeasible, the QMM modifications to electroweak processes might yield tiny corrections to flavor observables or \emph{rare-decay} branching ratios:
\begin{itemize}
    \item Neutral Meson Oscillations: Processes like \(K^0\leftrightarrow \overline{K}^0\) or \(B^0\leftrightarrow \overline{B}^0\) could display minuscule QMM-induced phase shifts if the local imprinting at each cell modifies the effective Hamiltonian.
    \item Lepton Flavor Violation: For example, \(\mu \to e\,\gamma\) or \(\mu\to e\, e^+ e^-\). Should QMM interactions cause slight mixing-angle or coupling changes, rare decays might shift by tiny amounts---still below current sensitivity, but potentially observable in future high-precision experiments.
    \item CP Violation: Local QMM imprint operators could alter CP phases in CKM or PMNS frameworks. Although stringent constraints already exist, subtle CP-violating effects could hint at QMM’s Planck-scale discretization.
\end{itemize}

These potential signals would be highly suppressed (e.g., by factors \(\sim (E/M_{\mathrm{P}})^n\)), but they remain an enticing avenue for probing Planck-scale unitarity. 

By extending QMM imprint mechanisms to SU(2)\(_\mathrm{L}\)\(\times\)U(1)\(_Y\), we have shown that electroweak symmetry breaking, the Higgs mechanism, and flavor mixing can be embedded within Planck-scale cells without violating gauge invariance. We next aim to \emph{unify} this formalism with the strong-interaction QMM approach (Section~\ref{sec:strong}), as well as with gravity, to produce a \emph{single} QMM Hamiltonian describing all known forces. That goal is taken up in Section~\ref{sec:unifiedHamiltonian}.

\section{Unified QMM Hamiltonian for Strong, Weak, and Gravitational Interactions}
\label{sec:unifiedHamiltonian}

Up to this point, we have shown how \emph{quantum imprints} in discretized space--time cells can encode information about gravity~\cite{neukart2024quantum}, electromagnetism, the strong interaction (SU(3)\(_\mathrm{c}\)), and the weak interaction (SU(2)\(_\mathrm{L}\)\(\times\)U(1)\(_Y\)). We now combine these ingredients into a single QMM-based Hamiltonian that unifies \emph{all} known interactions under a local, Planck-scale discretized framework. The resulting theory aims to preserve unitarity and locality for processes ranging from low-energy hadron formation to black hole evaporation, potentially offering a coherent quantum gravity and Standard Model (SM) description while also acknowledging foundational challenges such as Lorentz symmetry at \mbox{Planck scales.}

\subsection{Combining \texorpdfstring{SU(3)\(\times\)SU(2)\(\times\)U(1)\(\times\)\{Gravity\}}{SU(3)xSU(2)xU(1)x\{Gravity\}} in QMM}
\label{subsec:combining_all}

The SM gauge group is  \vspace{-10pt}
\[
\mathrm{SU}(3)_\mathrm{c} \;\times\; \mathrm{SU}(2)_\mathrm{L} \;\times\; \mathrm{U}(1)_Y,
\]
and gravity is incorporated via a discrete geometric structure at the Planck scale~\cite{rovelli1998loop, ashtekar2004background}. Within QMM, each Planck cell \(x\in\mathcal{X}\) thus carries:

\begin{itemize}
    \item \(\hat{G}_\mu^a(x)\), the \(\mathrm{SU}(3)\) color gluon fields (with indices \(a=1,\dots,8\)),
    \item \(\hat{W}_\mu^i(x)\), the \(\mathrm{SU}(2)_\mathrm{L}\) weak isospin fields (with \(i=1, 2, 3\)),
    \item \(\hat{B}_\mu(x)\), the \(\mathrm{U}(1)_Y\) hypercharge field,
    \item \(\hat{e}^a_\mu(x)\) or analogous geometric/gravitational operators (e.g., discrete tetrads, spin network edges)~\cite{rovelli1998loop, neukart2024quantum},
    \item matter fields \(\hat{\psi}(x)\) (quarks, leptons, Higgs) transforming under these gauge and gravitational sectors,
    \item QMM imprint operators \(\hat{I}_x^{(\mathrm{grav},\, \mathrm{QCD},\, \mathrm{EW})}\) responsible for storing and retrieving local quantum data.
\end{itemize}

In principle, each cell’s Hilbert space \(\mathcal{H}_x\) factors into subspaces for geometry, color, electroweak, and QMM memory degrees of freedom, though in practice these can be highly entangled. How such factorization recovers macroscopic Lorentz invariance or how the QMM memory modes fit into the cosmological energy budget are part of the broader open questions discussed in Section~\ref{sec:comparison}.

\subsection{Constructing the Total Hamiltonian}
\label{subsec:totalH_construction}

We write the \emph{unified} Hamiltonian in schematic form:
\begin{equation}\label{eq:total_unified_H}
    \hat{H} =
    \underbrace{\hat{H}_{\mathrm{grav}}}_{\substack{\text{discrete gravity}\\\text{(LQG-like)}}}
    + \underbrace{\hat{H}_{\mathrm{QCD}}}_{\substack{\text{SU(3) fields}\\\text{+ quarks}}}
    + \underbrace{\hat{H}_{\mathrm{EW}}}_{\substack{\text{SU(2)\(\times\)U(1)}\\\text{+ Higgs}}}
    + \underbrace{\hat{H}_{\mathrm{QMM}}}_{\substack{\text{intrinsic memory}\\\text{dynamics}}}
    + \underbrace{\hat{H}_{\mathrm{int}}}_{\substack{\text{QMM--field}\\\text{couplings}}}.
\end{equation}

\begin{figure}[H]

\includegraphics[width=1\textwidth]{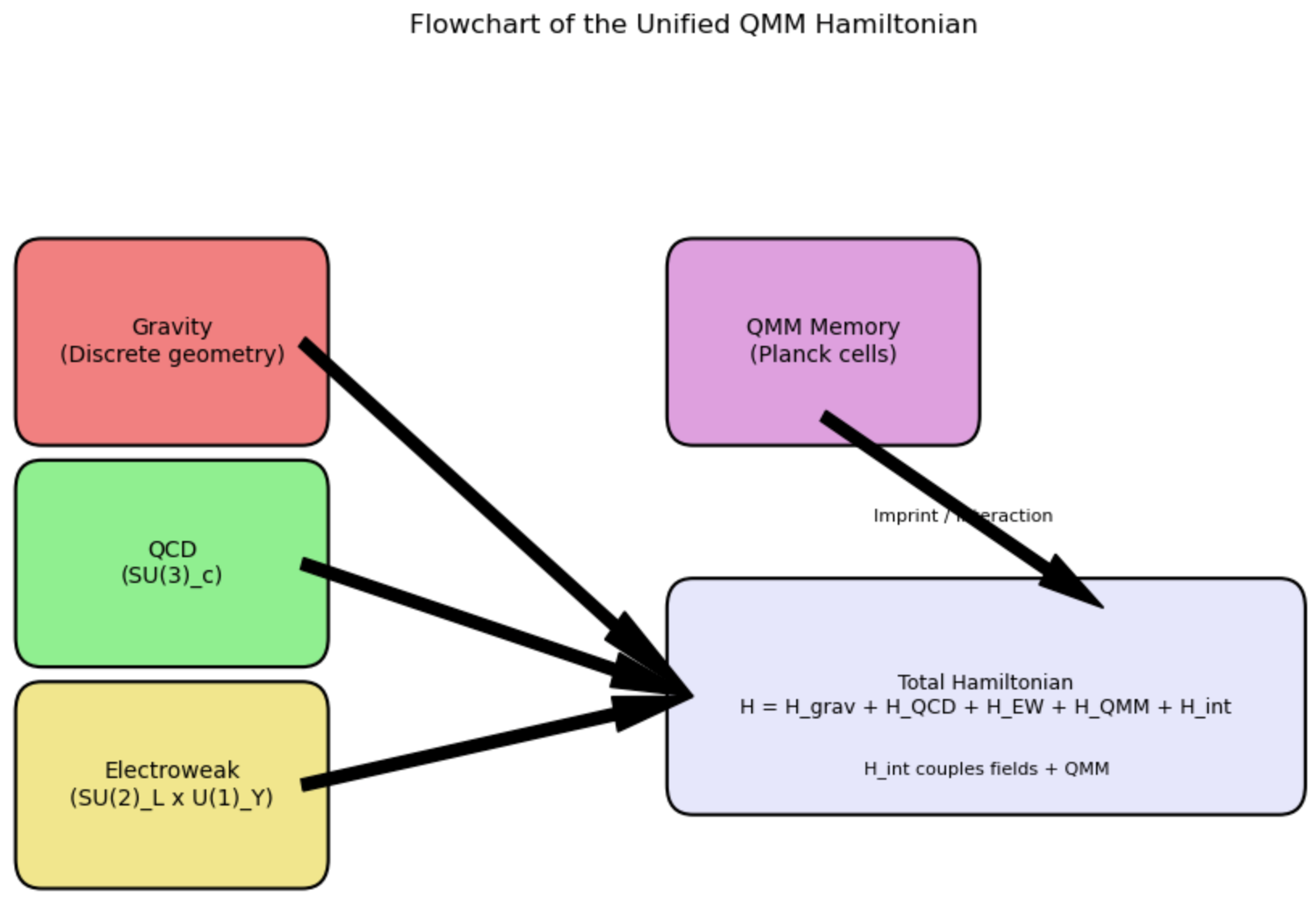}
\caption{Flowchart illustrating the construction of the unified QMM Hamiltonian. Each sector (\emph{Gravity}, \emph{QCD}, \emph{Electroweak}, and \emph{QMM Memory}) feeds into the total Hamiltonian \(\hat{H}\). The interaction Hamiltonian \(\hat{H}_{\mathrm{int}}\) couples the imprint degrees of freedom to the gauge and gravitational fields, ensuring local unitarity.}
\label{fig:unified_flowchart}
\end{figure}

A high-level flowchart of how each sector---Gravity, QCD, Electroweak, and QMM memory---combines into the total Hamiltonian \(\hat{H}\) is shown in Figure~\ref{fig:unified_flowchart}.
\vspace{-4.0pt}

\subsubsection{Gravitational Sector \(\hat{H}_{\mathrm{grav}}\)}
This portion encodes the dynamics of discrete geometry (e.g., spin network edges/nodes, or discrete tetrads)~\cite{rovelli1998loop, ashtekar2005}. Within QMM, gravity’s imprint operators capture local curvature or spin-connection data at Planck cells. Hamiltonian constraints generating diffeomorphisms or local Lorentz transformations remain valid at the discrete level, aiming to preserve general covariance in a continuum limit.

\subsubsection{QCD Sector \(\hat{H}_{\mathrm{QCD}}\)}
Here we include the SU(3) gauge fields \(\hat{G}_\mu^a\), quark fields \(\hat{\psi}_\mathrm{q}\), and potentially link variables \(\hat{U}_{x,\mu}^{(\mathrm{color})}\). Color imprint operators \(\hat{I}_x^{(\mathrm{color})}\) uphold local SU(3) invariance, capturing flux tubes and confinement phenomena~\cite{wilson1974, peskin1995introduction}.

\subsubsection{Electroweak Sector \(\hat{H}_{\mathrm{EW}}\)}
This sector consists of the SU(2)\(_\mathrm{L}\)\(\times\)U(1)\(_Y\) fields, the Higgs doublet, and leptons/quarks in their appropriate representations. After spontaneous symmetry breaking, the $W$, $Z$, and photon modes arise. QMM imprint operators \(\hat{I}_x^{(\mathrm{weak})}\) govern local isospin, hypercharge, and flavor transitions~\cite{weinberg1995quantum}.

\subsubsection{Intrinsic QMM Sector \(\hat{H}_{\mathrm{QMM}}\)}
Each cell has an internal memory subspace that can evolve independently even absent external fields, reflecting the “self-dynamics” of the QMM degrees of freedom~\cite{neukart2024quantum}. Such terms allow space--time quanta not to remain static but to reorder or propagate internal states, potentially underlying geometry's quantum fluctuations.

\subsubsection{Interaction Hamiltonian \(\hat{H}_{\mathrm{int}}\)}
Finally, the \emph{imprint} couplings arise here:
\[
   \hat{H}_{\mathrm{int}} \;=\; 
   \sum_{x \in \mathcal{X}} \Bigl[
     \hat{\Phi}_{\mathrm{fields}}(x) \,\otimes\, \hat{I}_x^\dagger
     + \hat{\Phi}_{\mathrm{fields}}^\dagger(x) \,\otimes\, \hat{I}_x
   \Bigr],
\]
where \(\hat{\Phi}_{\mathrm{fields}}(x)\) denotes any of the gravitational, color, or electroweak operators that locally imprint on the QMM cell. By design, \(\hat{H}\) is Hermitian, ensuring the total evolution \( \hat{U}(t) = e^{-i \hat{H} t} \) is unitary.

\subsection{Ensuring Local Gauge Invariance and Unitarity}
\label{subsec:gauge_invariance_unitarity}

\subsubsection{Local Gauge Transformations}
Under SU(3)\(\times\)SU(2)\(\times\)U(1), each cell’s gauge fields and matter fields transform as
\[
(\hat{G}_\mu^a, \hat{W}_\mu^i, \hat{B}_\mu) \;\;\mapsto\;\; 
(\hat{G}_\mu^{\prime a}, \hat{W}_\mu^{\prime i}, \hat{B}_\mu'),
\quad
\hat{\psi}(x) \;\;\mapsto\;\;\hat{\psi}'(x),
\]
where the transformations are parameterized by local color rotations, weak isospin rotations, and hypercharge phases. The imprint operators \(\hat{I}_x\) must remain covariant or invariant, preserving gauge symmetry in each Planck cell~\cite{wilson1974, peskin1995introduction}.

\subsubsection{Gravitational Covariance}
Similarly, local Lorentz or diffeomorphism invariance is reflected in how discrete gravitational variables and imprint operators transform. Technical details often borrow from loop quantum gravity or spin foam~\cite{rovelli1998loop, ashtekar2005}, but QMM adds an “information” dimension to each cell, storing curvature/torsion data as quantum states.

\subsubsection{Global Unitarity}
Because \(\hat{H}\) from Eq.~\eqref{eq:total_unified_H} is Hermitian, matter + gauge + QMM memory evolve unitarily as a closed system. No external boundary conditions or nonlocal couplings are needed to ensure conservation of information. Local imprinting at every cell ensures that each quantum event is recorded and, in principle, retrievable---even under black hole collapse.

\subsection{Natural \texorpdfstring{UV}{UV} Cutoff and Coupling Behaviors}
\label{subsec:UV_cutoff_couplings}

A recurring theme of quantum field theory is divergence at high energies. Continuum loop integrals can diverge unless renormalized; couplings may run to unphysical (e.g., Landau pole) values. QMM imposes a fundamental discretization at the Planck scale, capping momenta at \(p_{\mathrm{max}}\sim M_{\mathrm{P}}\). The dimension \(d_x\) of each cell’s Hilbert space bounds local field modes~\cite{hossenfelder2013minimal, neukart2024quantum}.

\subsubsection{Renormalization Flow at Planck Scale}
Below Planck energies, conventional RG flows for QCD or electroweak couplings apply. Approaching $M_{\mathrm{P}}$, the QMM structure halts further divergences. In principle, one might unify SM couplings with gravity in this discrete setting, such that couplings approach finite “fixed points” instead of diverging~\cite{weinberg1979ultraviolet, reuter1998average}.

\subsubsection{Interplay of Gravity and Gauge Fields}
Since gravity strengthens near $M_{\mathrm{P}}$, the QMM memory for geometry merges with gauge memory at those scales. This unification can alter naive QCD or electroweak RG flows, potentially influencing standard GUT scenarios or offering alternatives to unify. While the specifics remain open, QMM’s built-in cutoff plus local unitarity provides a framework to investigate.

\subsection{Summary of the Unified Hamiltonian Approach}
\label{subsec:unified_summary}

By assigning each Planck cell a finite-dimensional Hilbert space that accommodates gravitational and gauge imprint operators, we merge SU(3)\(\times\)SU(2)\(\times\)U(1) and gravity into a single quantum system. Key features include:
\begin{enumerate}
    \item Local Interactions at Each Cell: All imprinting occurs on-site, so no nonlocal or acausal effects arise.
    \item Global Unitarity: Hermiticity of $\hat{H}$ ensures the entire system—fields plus QMM memory—remains unitary.
    \item Finite UV Cutoff: Each cell’s dimensional limit imposes a Planck-scale cap on high-momentum modes, softening divergences.
    \item Recovery of SM Physics at Low Energies: At scales $\ll M_{\mathrm{P}}$, the usual SM and semiclassical gravity should emerge, matching observed phenomenology.
\end{enumerate}

With this unified Hamiltonian in hand, we now explore (Section~\ref{sec:applications}) how QMM applies to processes like black hole evaporation, baryogenesis, and other extreme regimes, suggesting potential observational tests of Planck-scale discreteness and unitarity.

\section{Applications and Implications}
\label{sec:applications}

With the QMM framework extended to include the full \(\mathrm{SU}(3)_\mathrm{c} \times \mathrm{SU}(2)_\mathrm{L} \times \mathrm{U}(1)_Y\) Standard Model gauge group (plus gravity), we can explore a range of physical scenarios. In this section, we focus on four major areas of interest: 
\begin{enumerate}
    \item Black hole evaporation with colored and electroweak-charged matter.
    \item Baryogenesis and sphaleron processes in the early universe.
    \item Non-perturbative and high-energy implications for confinement and \textls[-15] flavor physics.
    \item Possible observational prospects in cosmic rays, gravitational waves, and beyond.
\end{enumerate}

These applications highlight how QMM’s discretized, local-memory perspective offers new ways to address longstanding puzzles such as the black hole information paradox, baryon asymmetry, and Planck-scale physics.

\subsection{Black Hole Evaporation with Full Standard Model Content}
\label{subsec:bh_full_SM}

\subsubsection{Colored and Weakly Interacting Matter Infalling}
When matter carrying color charge (quarks/gluons) or electroweak charge (leptons, W/Z bosons) collapses into a black hole, the local interactions near the event horizon are governed by the combined imprint operators for QCD and electroweak fields (Sections~\ref{sec:strong} and \ref{sec:weak}). As these particles cross the horizon, \emph{quantum imprints} are deposited into the QMM cells at or near the horizon, encoding information about color, flavor, hypercharge, and isospin. This record-keeping extends earlier QMM treatments of neutral or purely electromagnetic matter~\cite{neukart2024quantum} to the broader SM sector.

\subsubsection{Non-Thermal Correlations in Hawking Radiation}
\begin{figure}[H]

\includegraphics[width=1\textwidth]{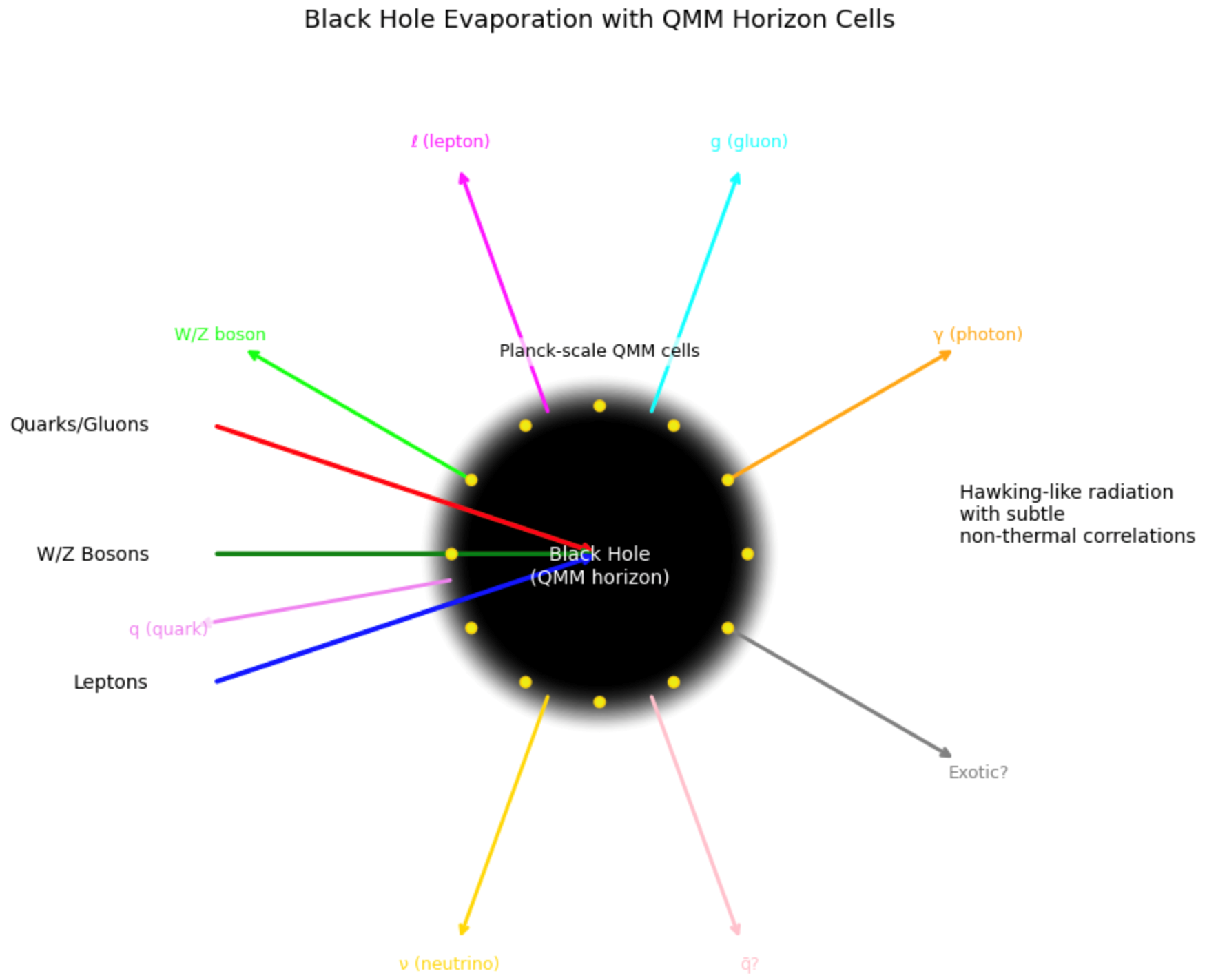}
\caption{Schematic view of a black hole formed from colored and electroweak-charged matter. As it evaporates, outgoing quanta (gluons, photons, leptons) interact with QMM horizon cells that store color/electroweak imprints, introducing subtle non-thermal correlations in the Hawking radiation.}
\label{fig:bh_evap_illustration}
\end{figure}

Hawking’s semiclassical calculation indicates that black hole radiation is approximately thermal, offering little information about the black hole interior~\cite{hawking1975, hawking1976}. Under QMM, however, the outgoing quanta---including gluons, quarks, photons, and leptons---undergo additional interactions with the horizon cells that store QCD and electroweak imprints. Over the course of evaporation, subtle \emph{color} and \emph{flavor} correlations become imprinted on the radiation. Although these deviations from pure thermality are small, they ensure global unitarity (Figure~\ref{fig:bh_evap_illustration}). In principle, after complete evaporation, the final state is entangled with QMM degrees of freedom in such a way that no net information is lost~\cite{preskill1992, unruh1976notes}.
\vspace{-3.0pt}

\paragraph{{Color Singlet Retrieval and the End State}}
In usual QCD processes, asymptotic color-neutral hadrons eventually form. In a black hole, any color “hair” is naively hidden behind the horizon. However, QMM memory for color flux implies that evaporating partons can retrieve color data in correlated jets, albeit in a Planck-scale or redshifted context. When the black hole nears Planckian mass, the gravitational imprint sector merges with QCD imprint states~\cite{rovelli1996black, neukart2024quantum}. The entire process stays unitary: the black hole does not disappear leaving hidden color microstates, since QMM degrees of freedom release color or flavor data back to the external field.

\medskip
 %MDPI: Please confirm if this paragraph' format is correct and should be kept. FN: yes, it is correct but we do not necessarily need the header

In parallel to QMM-based unitarity, other recent approaches propose that black holes might be hollow quantum shells with strong fluctuations~\cite{ShellModel2014,ShellModel2023,ShellModel2022,ShellModel2024}. From a QMM viewpoint, the horizon region---whether it is shell-like or a discrete horizon layer---is still endowed with local Planck-scale memory cells that record infalling data. Thus, one can see these “shell” models as complementary to the QMM mechanism, with both challenging the classical notion of a featureless horizon.

\subsection{Baryogenesis, Sphalerons, and Early-Universe Processes}
\label{subsec:baryogenesis_sphalerons}

\subsubsection{Electroweak Sphalerons in the QMM Framework}
Electroweak baryogenesis relies on \emph{sphaleron} transitions that violate baryon-plus-lepton number (\(B+L\)) while preserving \(B-L\). In continuum field theory, sphalerons are non-perturbative solutions to the SU(2)\(_\mathrm{L}\) gauge equations at high temperatures~\cite{weinberg1995quantum}. Under QMM, sphaleron configurations amount to large changes in local imprint operators for the \(\mathrm{SU}(2)\) field. As the Higgs field evolves during phase transitions, QMM’s finite memory tracks these transitions, potentially altering the net baryon asymmetry. While this does not alone solve the matter--antimatter asymmetry problem, it gives a discretized avenue for analyzing real-time sphaleron evolution near the Planck epoch, with fewer UV divergences than in continuum treatments.

\subsubsection{QCD Phase Transitions and Color Imprints}
The QCD sector also undergoes significant changes in the early universe (e.g., at temperatures around \(\Lambda_{\mathrm{QCD}}\)). Within QMM, color confinement emerges as flux tubes become imprinted in discrete cells (Section~\ref{sec:strong}). At sufficiently high temperatures, confinement can be transiently lost, but as the universe cools, hadronization reasserts itself. Because each cell has a limited capacity for storing color, one can in principle model how domain walls, topological defects, or other non-perturbative structures arise in a discrete manner. Though direct observational signatures may be faint, such processes could leave traces in gravitational wave backgrounds or other cosmological relics.

\subsection{Confinement Scale, Flavor Physics, and High-Energy Collisions}
\label{subsec:confinement_flavor}

\subsubsection{Hadronization in Extreme Energies}
Relativistic heavy-ion collisions (e.g., at the Large Hadron Collider) produce hot, dense quark--gluon plasmas. Although these energies are still far below $M_{\mathrm{P}}$, final-state correlations might reflect how QMM’s color memory influences color flux tubes or topological configurations. Subtle anomalies in jet substructure or flavor composition could, in principle, reveal Planck-scale discretization if extremely precise data were available. While likely much smaller than typical QCD effects, such signals remain intriguing for glimpses of new physics.

\subsubsection{Rare Decays and Flavor-Changing Processes}
As noted in Sections~\ref{sec:weak} and~\ref{subsec:rare_decays}, QMM effects could manifest as tiny deviations in flavor transitions (quark mixing, neutrino oscillations). Minute shifts in CP violation phases could alter processes like \(K^0\)--\(\overline{K}^0\) mixing or lepton-flavor-violating decays. Though beyond current experimental reach, next-generation precision experiments might observe systematic discrepancies if QMM imprint operators introduce corrections on the order of~\((E/M_\mathrm{P})^n\).

\subsection{Observational Prospects: Cosmic Rays, Gravitational Waves, {etc.}}
\label{subsec:observations_prospects}

\subsubsection{Cosmic Rays and Primordial Black Holes}
Primordial black holes (PBHs), if they formed in the early universe, may be evaporating today, emitting quarks, gluons, gauge bosons, and leptons~\cite{preskill1992}. QMM correlations in that radiation could yield non-thermal spectral features. If advanced cosmic-ray or gamma-ray observatories detect anomalies in PBH evaporation signals, such results may indirectly endorse the QMM picture of local unitarity~\cite{hawking1975, hawking1976}. Conversely, null results could constrain PBH populations or limit QMM’s parameter space.

\subsubsection{Gravitational Waves from Black Hole Mergers}
Recent gravitational wave detections have opened new vistas on black holes~\cite{abbott2016Observation}. Should QMM alter the ringdown or late-time merger behavior via horizon-scale memory interactions, small deviations in the waveform may appear. Although challenging to measure, next-generation detectors (Einstein Telescope, Cosmic Explorer) might detect slight phase shifts or echoes associated with QMM horizon cells.

\subsubsection{Analog Experiments and Quantum Simulators}
Finally, as previously suggested~\cite{neukart2024quantum}, analog experiments with “artificial horizons” (e.g., in fluid systems, optical waveguides, or superconducting circuits) might model aspects of local QMM imprinting. Similarly, discrete lattice gauge theories on quantum computing platforms could approximate QMM’s Planck-cell structure, offering indirect tests of the imprint-and-retrieval mechanism at lower energies. Although such analogs cannot replicate true Planck-scale physics, they may validate the internal consistency of the QMM approach.

\medskip
\noindent
In sum, %MDPI: Please confirm if this paragraph' format is correct and should be kept. FN: it is correct, as it summarizes a few key aspects of this section
 from black hole evaporation to early-universe baryogenesis and high-energy collider processes, the QMM approach furnishes a unifying lens through which to address puzzles like color confinement, neutrino mixing, and unitarity at horizons. These phenomena illustrate QMM’s promise for bridging quantum gravity and particle physics within a single coherent framework. In Section~\ref{sec:comparison}, we compare QMM with other unification attempts (holography, ER=EPR, loop quantum gravity, minimal-length theories), underscoring its distinct advantages of local discretization and information-centric unitarity.

\section{Comparison with Other Approaches}
\label{sec:comparison}

The QMM framework seeks to unify quantum mechanics and gravity, along with the full Standard Model gauge group, by positing that Planck-scale space--time cells act as local \emph{information reservoirs}. This stands in contrast to many traditional approaches to the black hole information paradox and quantum gravity. Below, we briefly compare QMM with four well-established lines of research: (1) holography and AdS/CFT, (2) ER=EPR and wormhole-based nonlocality, (3) loop quantum gravity and spin foam models, and (4) minimal length or causal set theories. These comparisons highlight the local, discretized, and gauge-invariant perspective QMM brings, as well as open points of synergy or divergence.

\subsection{Holography and the AdS/CFT Correspondence}
\label{subsec:holography_AdSCFT}

\subsubsection{Boundary vs.\ Bulk Encoding of Information}
Holographic approaches, including the AdS/CFT correspondence, assert that the degrees of freedom in a $(d+1)$-dimensional bulk gravitational system are equivalent to those of a $d$-dimensional boundary conformal field theory~\cite{maldacena1998, susskind1995, tHooft1993Dimensional}. Information about bulk black holes is “holographically” stored on the boundary, which can resolve the black hole information paradox by transforming it into a purely field-theoretic problem in \mbox{lower dimensions}.

\paragraph{QMM Contrast}
While holography relies on boundary degrees of freedom and often requires a specific asymptotic geometry (e.g., AdS), QMM encodes information \emph{locally} throughout the four-dimensional bulk. Planck-scale cells store quantum imprints of matter and gauge fields at each point in space--time, rather than on a boundary. This local encoding remains valid in asymptotically de Sitter or other cosmological spacetimes, where standard AdS/CFT techniques may not directly apply.

\subsubsection{Applicability to Realistic Cosmologies}
Since our universe appears to be close to de Sitter-like expansion (with a positive cosmological constant), AdS-based holography is not obviously the final word on real-world quantum gravity. QMM offers a geometrically local scenario that may be extended to various backgrounds, preserving unitarity while avoiding reliance on a conformal~boundary.

\subsection{ER=EPR Conjecture and Wormhole-Based Nonlocality}
\label{subsec:ER_EPR}

\subsubsection{Einstein--Rosen Bridges as Entanglement}
The ER=EPR proposal~\cite{maldacena2013cool} posits that the nonlocal entanglement of quantum mechanics (EPR pairs) can be interpreted geometrically via Einstein--Rosen bridges (wormholes). Some black hole information paradox resolutions lean on this topology-changing interpretation, suggesting that information retrieval is facilitated by hidden wormholes connecting interior and exterior degrees of freedom.

\paragraph{QMM Contrast}
QMM does not assume or require large-scale wormholes or topological changes to maintain unitarity. Instead, local interactions at the Planck scale encode and later retrieve quantum information from the horizon cells—no nonlocal bridging is necessary. This maintains classical large-scale space--time topology, respecting the equivalence principle and causality at macroscopic distances.

\subsubsection{Observer Independence}
Wormhole-based solutions often introduce subtleties about observer dependence and potential horizons merged by nontrivial topology. QMM, by contrast, implements an \emph{observer-independent} local memory: each Planck cell’s Hilbert space evolves unitarily regardless of who observes it, sidestepping issues of observer-dependent complementarity or firewall paradoxes.  

\subsection{Loop Quantum Gravity, Spin Foams, and Causal Sets}
\label{subsec:LQG_spinfoam}

\subsubsection{Discrete Geometry Emphasis}
Loop quantum gravity (LQG) and spin foam models propose that space--time itself is composed of discrete spin networks, where edges carry quantized areas and volumes~\cite{rovelli1998loop, ashtekar2004background}. Similarly, causal set theory posits that the fundamental structure is a partially ordered set of events consistent with relativistic causality~\cite{bombelli1987spaceTime}. In each, the continuum emerges at large scales, but the geometry is fundamentally discrete.

\paragraph{QMM vs.\ LQG/Spin Foam}
QMM shares the discrete flavor of these approaches, but it \emph{enhances} the notion of a discretized geometry with explicit \emph{information-storage} capacity. Each Planck cell in QMM is not merely a geometric datum (area, spin), but also a finite-dimensional Hilbert space that can store local quantum states. Where LQG focuses on geometry quantization, QMM focuses on \emph{information} quantization. The two ideas could be complementary: one might combine QMM’s local memory concept with LQG’s well-studied discrete geometry to ensure that matter, gauge, and gravitational interactions are all captured consistently.

\subsubsection{Non-Abelian Gauge Fields and Lattice-Like Formulations}
Although LQG and spin foam models typically focus on gravity, they can in principle incorporate gauge fields. QMM uses a lattice-like approach to embed gauge fields in Planck cells. Causal set theory does something analogous with partial orders. Thus, synergy may arise if one treats geometry with LQG/spin foams while letting QMM handle the “information imprint” aspect. The union could provide a robust discrete quantum \mbox{gravity + gauge} theory picture.

\subsection{Minimal Length Theories and Causal Set Ideas}
\label{subsec:minimal_length}

\subsubsection{UV Regularization and Planck-Scale Cutoffs}
Various scenarios suggest the existence of a minimal length or a fundamental UV cutoff at or near the Planck scale to tame divergences in quantum field theory~\cite{hossenfelder2013minimal}. Causal set theory also postulates that spacetime is fundamentally discrete, with each event part of a partially ordered set respecting causality~\cite{bombelli1987spaceTime}.

\paragraph{QMM Perspective}
QMM indeed posits a minimal length scale—each cell has finite volume $\sim$$l_{\mathrm{P}}^3$. Moreover, each cell is associated with a finite-dimensional Hilbert space, capping the number of local degrees of freedom. Hence, QMM can be viewed as a \emph{physical} realization of minimal length ideas: beyond imposing a cutoff, QMM treats the cell as an active memory site. This helps unify the notion of discrete spacetime with a mechanism for storing quantum information, maintaining unitarity in processes like black hole evaporation.

\subsubsection{Cause--Effect and Local Encoding}
Causal set approaches ensure that no event has more than one immediate predecessor or successor outside its light cone. QMM ensures strict locality—imprinting interactions only occur within a single Planck cell or between neighboring cells. While not identical, both reflect the principle that fundamental causality is essential to quantum gravity.

\subsection{Summary of Comparisons}
\label{subsec:summary_comparisons}

\subsubsection{Local vs.\ Nonlocal Mechanisms}
Holography and ER=EPR typically involve nonlocal mappings (bulk--boundary correspondences, wormhole entanglement) to preserve unitarity. QMM keeps all interactions local at the Planck scale: each cell is an autonomous memory site, storing and releasing data strictly within its boundaries.

\subsubsection{Boundary vs.\ Bulk Encoding}
Whereas holography uses boundary degrees of freedom to encode bulk states, QMM organizes the bulk itself into discrete Planck cells. This may offer a more direct route to describing real cosmologies without specialized AdS boundaries.

\subsubsection{Geometry-Focused vs.\ Information-Focused Approaches}
Loop quantum gravity and spin foam models quantize geometry, while QMM extends that quantization with an information-centric viewpoint. These could be complementary rather than competing. The same goes for causal set theories, which emphasize partial orders but do not necessarily include an internal memory interpretation.

\subsection{Concluding Remarks on Comparisons}

In essence, QMM shares many goals with these other frameworks—resolving the black hole information paradox, unifying quantum field theory and gravity, taming UV divergences—yet differs in its \emph{local, discretized, memory-based} formalism. The next section addresses open challenges and technical issues in fully realizing QMM, from building precise non-Abelian imprint operators to integrating renormalization group flows and exploring experimental feasibility. 

\section{Challenges and Open Questions}
\label{sec:challenges}

Throughout this work, we have shown how the QMM framework can be extended to accommodate the full set of Standard Model interactions—namely SU(3)\(_\mathrm{c}\) for QCD and SU(2)\(_\mathrm{L}\)\(\times\)U(1)\(_Y\) for the electroweak sector—together with a discretized model of quantum gravity. Despite this progress, several conceptual and technical challenges remain to \mbox{be addressed:}

\begin{enumerate}
\item Exact Construction of Non-Abelian Imprint Operators: While we have sketched how to incorporate local SU(3)\(_\mathrm{c}\) and SU(2)\(_\mathrm{L}\) gauge invariance, a fully rigorous definition of imprint operators for the full range of non-Abelian interactions remains an open problem. Ensuring these operators transform properly (covariantly or as singlets) under local gauge transformations in a discrete lattice-like setup requires further elaboration, especially if we wish to handle strong self-interactions (e.g., gluon self-couplings) and mixed flavor transitions. Moreover, some proposed “shell-based” black hole models~\cite{ShellModel2014,ShellModel2023,ShellModel2022,ShellModel2024} might demand even more refined imprint definitions, should one try to merge QMM with such horizons.

\item Renormalization Group Analysis and Low-Energy Effective Theories: In conventional quantum field theory, coupling constants run with energy scale. In QMM, the finite-dimensional nature of each cell imposes a natural Planck-scale cutoff, but how this meshes with the usual renormalization group flow is not yet fully elucidated. Analyses must clarify if—and how—the discrete QMM structure modifies coupling unification scenarios or prevents Landau poles in a manner consistent with low-energy observations. Additionally, investigating how discrete Planck-cell memory degrees of freedom could alter the running of SM couplings near $M_{\mathrm{P}}$ remains an intriguing but complex endeavor.

\item Dynamical Quantum Geometry and Loop-Like Structures: While QMM provides a discretized scheme for storing quantum information, one typically wishes to see how discrete geometric variables themselves evolve. This raises questions about linking QMM with loop quantum gravity, spin foam models, or other discrete gravitational formalisms, so that the geometry of each cell is not just an external label but a dynamic entity influencing (and being influenced by) the imprint operators. Clarifying whether QMM’s “memory capacity” for spacetime curvature can reproduce or refine known results in loop quantum gravity is an active area for future study.

\item Quantum States of Horizon Cells and Near-Horizon Physics: Although black hole horizons offer the most striking application of QMM’s memory mechanism, a deeper treatment of the near-horizon degrees of freedom is needed. In particular, how exactly the horizon cells entangle with outgoing modes, and how one might observe non-thermal signatures in practice, remains a major challenge—both theoretically and experimentally. This difficulty is compounded by proposals that black holes are quantum shells~\cite{ShellModel2014,ShellModel2023,ShellModel2022,ShellModel2024}, potentially requiring the QMM horizon layer to be interpreted in a shell-like context.

\item Experimental Feasibility and Observational Strategies: Direct tests of Planck-scale phenomena lie beyond current technology. Nevertheless, indirect signals—such as small deviations in rare decay processes, flavor transitions, or possible anomalies in gravitational wave ringdown—may provide windows into QMM effects. Identifying robust, model-specific predictions that could be probed by next-generation experiments is an ongoing task. Additionally, clarifying potential impacts on cosmic-ray or gamma-ray spectra from primordial black holes may shed indirect light on QMM-induced non-thermal correlations.

\item Computational and Numerical Simulations: Simulating lattice gauge theories coupled with a discrete “memory matrix” is non-trivial, particularly if one aims to include dynamical geometry. Efforts to develop efficient classical or quantum algorithms to explore QMM’s predictions—e.g., for black hole microstates or baryogenesis processes—would help quantify the theory’s viability and link it to real-world data. Investigations of 3D or 4D toy models, as well as quantum simulation platforms, could provide valuable proof-of-\mbox{concept~demonstrations}.
\end{enumerate}

These challenges outline the terrain ahead for extending and testing the QMM approach. Their resolution will require coordinated efforts, combining methods from lattice gauge theory, loop quantum gravity, quantum information science, and high-energy phenomenology to bring QMM from a conceptual framework toward a more concrete and falsifiable theory. Questions about the overall \emph{energetics} of Planck-cell memory and the restoration of effective Lorentz invariance at large scales remain especially pivotal for validating QMM’s physical consistency.

\section{Conclusion}\label{sec:conclusion}

By extending the Quantum Memory Matrix (QMM) framework—originally designed to reconcile quantum mechanics with gravity through discretized space--time cells—to encompass the \emph{entire} Standard Model gauge sector, we have taken a crucial step toward a unified picture of quantum fields and geometry. In previous publications, QMM was successfully integrated with gravitational degrees of freedom and electromagnetism, demonstrating how local ``quantum imprints'' in Planck-scale cells could preserve unitarity and record field interactions without resorting to nonlocal mechanisms or exotic geometries. This work completes the triptych by embedding the strong (SU(3)\(_\mathrm{c}\)) and electroweak (SU(2)\(_\mathrm{L}\)\(\times\)U(1)\(_Y\)) sectors into the QMM, thereby tying together all four fundamental interactions under a single discrete, memory-based hypothesis. We have shown that each cell can store color and electroweak charges in gauge-invariant imprint operators, unifying them with gravitational and electromagnetic degrees of freedom via a single Hamiltonian construction. Such a framework enables local unitarity even under black hole formation and evaporation while simultaneously imposing a natural ultraviolet cutoff by limiting the dimension of each cell’s Hilbert space.

Although open questions persist—ranging from refining non-Abelian imprint operators to analyzing renormalization flows at trans-Planckian energies—our results highlight the versatility and consistency of QMM as a \emph{fully unified} approach. In particular, we have identified possible connections to established quantum gravity programs (loop quantum gravity, causal sets) and provided toy simulation ideas that illustrate how local memory capacities might manifest in lattice or analog systems. These developments go beyond earlier achievements in QMM-plus-gravity or QMM-plus-electromagnetism, offering concrete pathways for future theoretical and numerical exploration.

By demonstrating how discretized Planck cells can serve as a universal storage mechanism for gravitational, strong, and electroweak fields alike, this paper underscores the promise of QMM to bridge high-energy physics and quantum gravity in a single, coherent framework. We therefore hope that numerical simulations, astrophysical signatures, and analog experiments continue to evolve around these principles, ultimately testing whether the QMM indeed paves a robust path toward a complete unification of \mbox{fundamental physics.}

\vspace{6pt}

%\supplementary{The following supporting information can be downloaded at: %MDPI: We found that Supplementary file is uploaded in our system. Please confirm if this Supplementary is necessary. If necessary, please check the accuracy of contents add its citation in main text. FN: this is the code reproducing the figures. It is not strictly necessary, but readers may be interested in how these were created. Please feel free to decide if Entropy deems it relevant to make it available. We do not have a preference, but thought it more of a convenience for readers.

%The following supporting information can be downloaded at: \linksupplementary{s1}. 
% }
 
\authorcontributions{%
F.N.\ conceptualized the research and drafted the outline; 
E.M.\ and V.V.\ helped shape the formalism; 
All authors reviewed, edited, and approved the final version. All authors have read and agreed to the published version of the manuscript. %MDPI: We added the standard note. Please confirm. FN: confirmed
}

\funding{This research received no external funding.}

%\institutionalreview{\hl{ } %MDPI: In this section, you should add the Institutional Review Board Statement and approval number, if relevant to your study. You might choose to exclude this statement if the study did not require ethical approval. Please note that the Editorial Office might ask you for further information. Please add “The study was conducted in accordance with the Declaration of Helsinki, and approved by the Institutional Review Board (or Ethics Committee) of NAME OF INSTITUTE (protocol code XXX and date of approval).” for studies involving humans. OR “The animal study protocol was approved by the Institutional Review Board (or Ethics Committee) of NAME OF INSTITUTE (protocol code XXX and date of approval).” for studies involving animals. OR “Ethical review and approval were waived for this study due to REASON (please provide a detailed justification).” OR “Not applicable” for studies not involving humans or animals. FN: not relevant to our study
%}

\dataavailability{No new data were created or analyzed in this study.}

\conflictsofinterest{%
The authors %MDPI: Please check whether any potential commercial interests are declared according to the relevant guidelines, which can be found here: https://www.mdpi.com/ethics#_bookmark17. FN: confirmed, no commercial or other interest conflicts
 declare no conflict of interest.
}

%%%%%%%%%%%%%%%%%%%%%%%%%%%%%%%%%%%%%%%%%%
\begin{adjustwidth}{-\extralength}{0cm}

\reftitle{References}

% Please provide either the correct journal abbreviation (e.g., according to the “List of Title Word Abbreviations” http://www.issn.org/services/online-services/access-to-the-ltwa/) or the full name of the journal.

%\bibliography{your_external_BibTeX_file}

%=====================================
% References, variant B: internal bibliography
%=====================================

%%%%%%%%%%%%%%%%%%%%%%%%%%%%%%%%%%%%%%%%%%
%% for journal Sci
%\reviewreports{\\
%Reviewer 1 comments and authors’ response\\
%Reviewer 2 comments and authors’ response\\
%Reviewer 3 comments and authors’ response
%}
%%%%%%%%%%%%%%%%%%%%%%%%%%%%%%%%%%%%%%%%%%
\PublishersNote{}
\end{adjustwidth}
\end{document}